\documentclass{svproc}
\usepackage[utf8]{inputenc}  
\usepackage{url}
\usepackage{graphicx}
\usepackage{subcaption} 

\usepackage{multirow}
\usepackage[xcdraw]{xcolor}

\usepackage{listings}
\begin{document}
\mainmatter              
\title{Bitcoin, a DAO?}
\titlerunning{Bitcoin, a DAO?}  
%
\author{Mark C. Ballandies 
 \and Guangyao Li\and Claudio J. Tessone
 }
 \authorrunning{Ballandies et al.}
 %
 \institute{University of Zurich, Zurich, Switzerland 
 }

\tocauthor{TBD Authors}
\institute{University of Zurich, Zurich, Switzerland \\
\email{markchristopher.ballandies@uzh.ch}}

\maketitle              

\begin{abstract}
 This paper investigates whether Bitcoin can be regarded as a decentralized autonomous organization (DAO), what insights it may offer for the broader DAO ecosystem, and how Bitcoin governance can be improved. First, a quantitative literature analysis reveals that Bitcoin is increasingly overlooked in DAO research, even though early works often classified it as a DAO. Next, the paper applies a DAO viability framework—centering on collective intelligence, digital democracy, and adaptation—to examine Bitcoin’s organizational and governance mechanisms. Findings suggest that Bitcoin instantitates key DAO principles by enabling open participation, and employing decentralized decision-making through Bitcoin Improvement Proposals (BIPs), miner signaling, and user-activated soft forks. However, this governance carries potential risks, including reduced clarity on who truly ‘votes’ due to the concentration of economic power among large stakeholders. The paper concludes by highlighting opportunities to refine Bitcoin’s deliberation process and reflecting on broader implications for DAO design, such as the absence of a legal entity. In doing so, it underscores Bitcoin’s continued relevance as an archetype for decentralized governance, offering important findings for future DAO implementations.
\end{abstract}
\section{Introduction}
\label{sec:dimensions}
Decentralization is on the rise in society \cite{Welling2023,yang2023designing,hoda2016multi,hoda2010organizing,laloux2014reinventing,wyrzykowska2019teal,Majumdar2023,helbing2023democracy,ballandies2023taxonomy} and blockchain-based
decentralized autonomous organizations (DAOs) accelerate this broader shift \cite{ballandies2024daos,spychiger2023organizing,hunhevicz2024decentralized}. Governed democratically through participatory algorithms, they transcend geographical, cultural, and traditional boundaries, often engaging thousands of members worldwide \cite{wright2021rise,Pournaras2020}. By removing clear distinctions between internal and external stakeholders, DAOs allow anyone to quickly influence decision-making and contribute to tasks \cite{spychiger2023organizing}.

DAOs have demonstrated their effectiveness through the management of substantial assets—sometimes exceeding \$500 million—and rapid capital deployment \cite{wright2021rise}. In addition to pooling capital, they streamline operations via algorithmic systems and blockchain tools, resulting in more efficient and cost-effective voting mechanisms \cite{wright2021rise,Pournaras2020}. Regular, ongoing voting replaces traditional periodic voting, making active participation more feasible \cite{wright2021rise}. Also, smart contracts can further facilitate liquid democracy, reducing complexity and costs in proxy-based voting while enhancing transparency and lowering the risk of fraud~\cite{wright2021rise}. 
Nonetheless, DAOs face notable challenges \cite{ballandies2024daos}, including declining participation \cite{wright2021rise,rikken2019governance,faqir2021comparative}, increasing centralization \cite{santana2022blockchain}, adaptation difficulties \cite{santana2022blockchain}, and balancing potentially conflicting values \cite{hunhevicz2024decentralized}.

As the first blockchain, Bitcoin demonstrated the potential of a decentralized network of peers operating without a central authority or legal entity. Several definitions have since been proposed for what constitutes a DAO (see Section \ref{sec:bitcoin_analysis}), with some viewing Bitcoin as a DAO and others not \cite{parmindar2025conceptual}.
This paper argues that Bitcoin, when positioned within the DAO viability framework (Table \ref{tab:dao_framework}), can be considered a DAO. Recognizing Bitcoin in this way would help researchers address the above mentioned challenges facing DAOs, as Bitcoin’s distinctive approach to digital democracy and adaptation—rooted in permissionlessness and transparency—broadens the design space for DAO development. In particular, the limited ability of many DAOs to adapt to changing conditions \cite{ballandies2024daos} could be addressed, which Bitcoin tackles through its focus on individual autonomy (Table \ref{tab:dao_framework}).

Hence, given Bitcoin’s success, we encourage and, with this work, provide the foundations for researchers to examine the advantages and disadvantages of its governance and organizational structure to derive insights for other DAOs. And for Bitcoin practitioners, this work offers a foundation for conceptual discussions on improving Bitcoin’s governance, a topic that continues to resurface within the community.
In particular, the following contributions are made in this work:
\begin{itemize}
    \item Quantitative Analysis of related work and their sentiment towards Bitcoin being a DAO, illustrating a change in perception over time (Section \ref{sec:background}).
    \item Analysis of Bitcoins governance and organizational mechanism via the DAO viability framework \cite{ballandies2024daos}, illustrating Bitcoins unique approach to Digital Democracy by using Bitcoin Improvement Proposals (BIP) for deliberation as well as hash power signaling and code forking for decision-making.
    \item Discussion of important design decisions in bitcoin, such as the lack of a legal entity or the permissionless entry into the organization, put into the context of recent developments, e.g. Bitcoin ETFs.
    \item Proposals to improve Bitcoins governance, such as its deliberation mechanism.
\end{itemize}

In Section \ref{sec:background} we quantitatively analyze related work on DAOs and identify their sentiment towards Bitcoin and provide an introduction to the DAO viability framework. We then describe Bitcoins organizational mechanisms in Section \ref{sec:bitcoin_main_section} and analyze them by applying the DAO viability framework. This is followed by a discussion, amongst others informing improvements to the Bitcoin governance mechanisms (Section \ref{sec:discussion}). In Section \ref{sec:conclusion} we conclude and give an outlook on future work. 

\section{Background}
\label{sec:background}
\subsection{Bitcoin, a DAO? A quantitative analysis}
\label{sec:bitcoin_analysis}

\begin{figure}[htbp]
    \centering
    \includegraphics[width=0.8\textwidth]{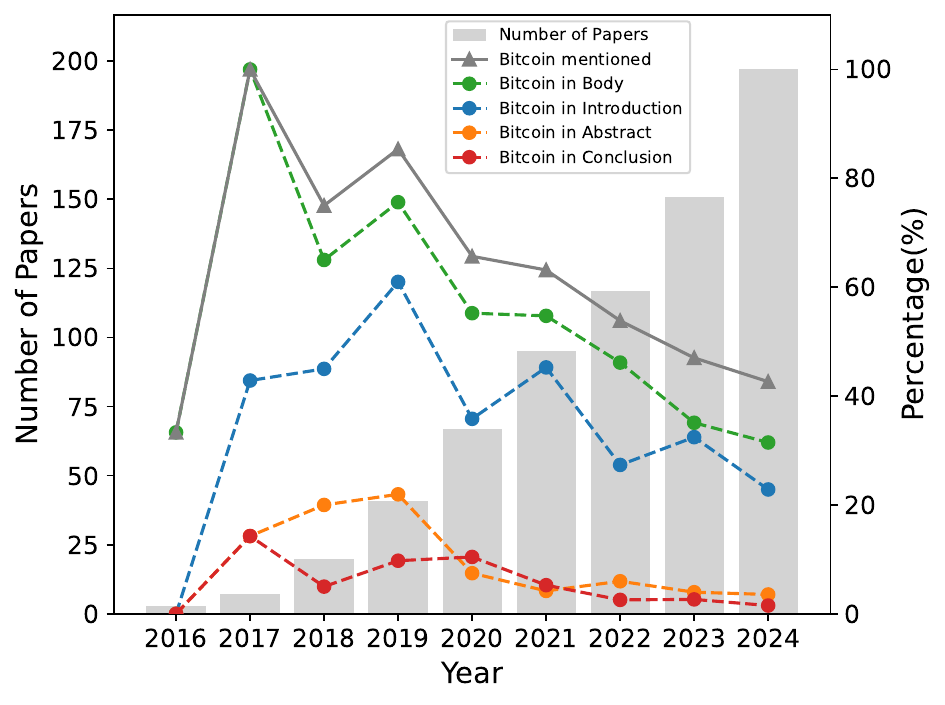} 
    \caption{Total numbers of DAO papers published (bar chart), the proportion of Bitcoin being mentioned in those papers (solid line), and the proportion of papers mentioning bitcoin in different parts of the paper (dashed lines).}
    \label{fig:bitcoin}
\end{figure}

As early as 2013, leading blockchain practitioners recognized Bitcoin as an example of a decentralized corporation \cite{buterin2013bootstrapping}, and it has since been identified in academic works as a DAO \cite{hsieh2018bitcoin}. In particular, Bitcoin depicts characteristics of an organization: Bitcoin is a multi-agent system, with established boundaries, a clear purpose, and coordinated actions \cite{hsieh2018bitcoin}. Notably, it even operates with a kind of marketing department \cite{buterin2013bootstrapping}, aligning with characteristics typical of an \textit{organization} \cite{puranam2014s}. Unlike traditional organizations though, Bitcoin is distinctive for its \textit{decentralized} governance, emerging from the complex interdependencies among its diverse stakeholders~\cite{fish2024AnalyzingBitcoin}. Also, its permissionless structure enables individuals to independently act on behalf of Bitcoin, ensuring its \textit{autonomy} which is backed and mediated by a set of self-executing rules on-chain (e.g. Nakamoto Consensus). This aligns Bitcoin to be classified as a DAO according to the definition given by Hassan and De Filippi~\cite{hassan2021decentralized}.

Nevertheless, a common definition of DAOs often restricts them to entities that explicitly use smart contracts for decision-making or manage collective treasuries, thus excluding Bitcoin as a DAO \cite{parmindar2025conceptual}. This view is gaining traction within the DAO research community, as shown in Figure \ref{fig:bitcoin}. The figure displays the number of papers indexed in Elseviers ScienceDirect database found with the search terms ("Decentralized Autonomous Organizations" OR DAO) AND (Blockchain OR Distributed Ledger). Over time, there has been a marked increase in DAO-related publications (grey bars in Figure \ref{fig:bitcoin}), in total 826 works have been published. However, an analysis of Bitcoin mentions within these papers reveals a decreasing trend in discussions of Bitcoin, e.g. in 2017 all papers discussed Bitcoin, whereas in 2024 this number decreased to $42\%$. Given Bitcoin's ongoing success, this decline suggests it is increasingly viewed as something other than a DAO, as its mechanisms would otherwise be more frequently examined in that context.
Analyzing this further, we used ChatGPT on the articles that mention Bitcoin to identify if the works consider Bitcoin as a DAO (Figure \ref{fig:bitcoin3}a). For this, we used the prompt as illustrated in Appendix \ref{sec:chatgp_prompt}. 
Over time, only 22 papers identify Bitcoin as a DAO, representing $6\%$ of papers discussing Bitcoin in the context of a DAO and $3\%$ of all papers.

\begin{figure}[htbp]
    \centering
    \begin{subfigure}{0.49\textwidth}
        \centering
        \includegraphics[width=\textwidth]{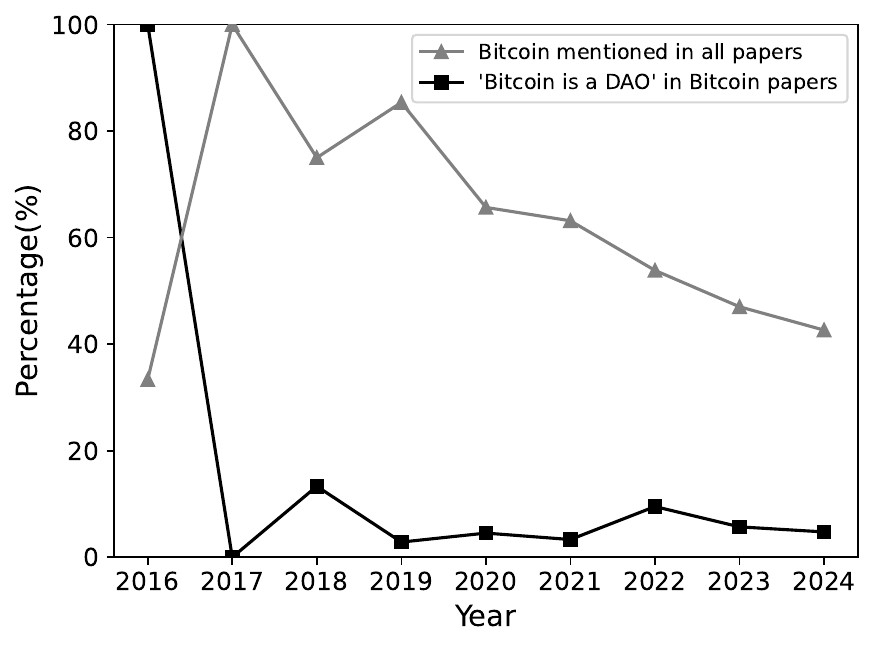}
        \label{fig:bitcoin_dao}
    \end{subfigure}
    \hfill
    \begin{subfigure}{0.49\textwidth}
        \centering
       
        \includegraphics[width=\textwidth]{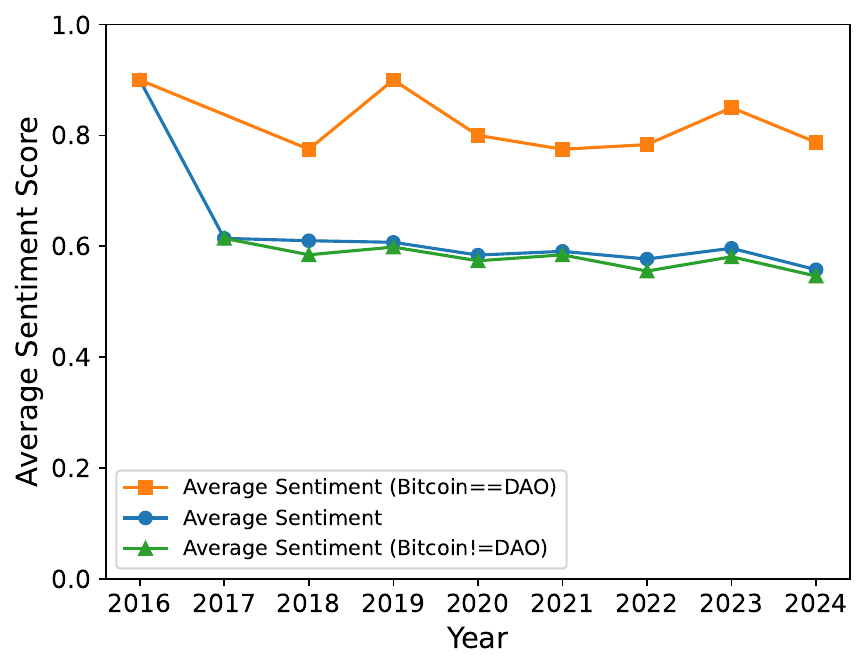}
         \label{fig:bitcoin2}
    \end{subfigure}
    \caption{\textbf{Left:} Proportion of academic papers discussing Bitcoin, and within that subset, the proportion considering Bitcoin as a DAO.
    \textbf{Right:} Average sentiment toward the statement "Bitcoin has great potential" across papers discussing Bitcoin, comparing those that explicitly consider Bitcoin as a DAO (Bitcoin == DAO) and those that do not (Bitcoin == False).}
    \label{fig:bitcoin3}
\end{figure}

Figure \ref{fig:bitcoin3}b presents the average sentiment of authors toward the statement "Bitcoin has great potential" in papers discussing Bitcoin as identified by ChatGPT. The average sentiment is approximately $60\%$ in these papers, rising to $80\%$ among those considering Bitcoin a Decentralized Autonomous Organization (DAO), suggesting a bias in this subset. This indicates a need for more critical examination of Bitcoin's governance and organizational mechanisms.

\subsubsection{Summary:} Given Bitcoin's dominant market position as well as governance challenges, the previous findings are concerning:
i) Despite Bitcoin's alignment with DAO characteristics, few researchers examine it through this lens. This oversight hinders the identification of resilient and efficient DAO mechanisms. ii)~Conversely, those who view Bitcoin as a DAO exhibit biases, impeding objective governance improvements. 

Therefore, more systematic investigations of Bitcoin as a DAO are essential for advancing DAO research and strengthening Bitcoin’s organizational resilience.


\subsection{DAO viability framework}

The DAO viability framework \cite{ballandies2024daos} is rooted in Complexity Science and Digital Democracy literature and defines DAOs through three key concepts: decentralization (a broad community of peers), autonomy (independence from coercive forces), and organization (mechanisms for coordination and decision-making). 
Each of these concepts facilitates a self-organization mechanism that can support the well functioning of a DAO: Collective intelligence is associated with the decentralizaton of the DAO and draws on the observation that decentralized networks can generate superior problem-solving capabilities when diverse participants contribute freely and transparently \cite{helbing2023democracy}.
Digital democracy is associated with the organizational aspect of a DAO and illustrates deliberative processes that produce legitimate decisions, exemplifying democratic innovations like liquid democracy and public reasoning frameworks \cite{yang2023designing,yang2025bridging}.
Adaptation is associated with the autonomy of the DAO and refers to the ability of complex systems to efficiently adjust to their environment
without centralized coordination or control \cite{heylighen2001science,kephart2003vision}. 

 These self-organization mechanisms require different principles do be manifested in DAOs \cite{ballandies2024daos}. Collective intelligence requires openness, transparency, privacy and free expression within the community. Digital democracy involves structured deliberation and fair voting mechanisms to distill and select solutions. Adaptation depends on community members to act autonomously, informed by feedback mechanisms, to implement decisions and respond to environmental changes.

Challenges in DAO design arise from the inadequate implementation of these principles \cite{ballandies2024daos}. Failures in adaptation can lead to inaction, while issues in digital democracy may result in poor decision-making. Additionally, insufficient collective intelligence can skew the identification of viable solutions. The DAO viability framework supports evaluating DAOs by analyzing their alignment with these mechanisms and their principles, providing insights for improving DAO governance and functionality, as demonstrated in Table \ref{tab:dao_framework} for Bitcoin. We clarify that these mechanisms and their principles are not merely normative ideals but serve as analytical criteria: each is grounded in operational principles (see Table \ref{tab:dao_framework}), and their assessment in Bitcoin’s case is based on observable mechanisms such as BIP discussions, open participation, and decentralized node signaling.


\subsection{Summary}
While Bitcoin exhibits characteristics of a DAO, its classification as such is increasingly debated. The DAO viability framework assesses a DAO's capacity to adapt efficiently to environmental changes. By applying this framework in the following onto Bitcoin, we show that Bitcoin can be seen as a DAO. This analysis critically examines Bitcoin's strengths and weaknesses, proposing potential improvements. Given Bitcoin's success, this approach may identify resilient and efficient mechanisms for DAO instantiation.


\begin{table}[]
\caption{Bitcoins accomplishment of the eight DAO viability principles \cite{ballandies2024daos} (last column). }
\begin{tabular}{llll} \hline
\textbf{}                                                                            & \textbf{Princip.}                                        & \textbf{Goal}                                                                                                                                                            & \textbf{\begin{tabular}[c]{@{}l@{}}Bitcoin \\ Accomp \\ lishment\end{tabular}} \\  \hline          
                                                                                     & Openess                                                   & Permeable borders for entering and leaving the organization.                                                                                                             & \begin{tabular}[c]{@{}l@{}}Medium-\\ High\end{tabular} \\
                                                                                     & Transp.                                            & \begin{tabular}[c]{@{}l@{}}Information related to the organization can be accessed by\\ all community members.\end{tabular}                                              & High                                                   \\
                                                                                     & Privacy                                                   & \begin{tabular}[c]{@{}l@{}}Community members can independently/ unmanipulated\\ explore and experiment with solutions.\end{tabular}                                      & \begin{tabular}[c]{@{}l@{}}Medium-\\ High\end{tabular}                             \\
\parbox[t]{2mm}{\multirow{-7}{*}{\rotatebox[origin=c]{90}{\multirow{2}{*}{\textbf{\begin{tabular}[c]{@{}l@{}}Collective\\ Intelligence\end{tabular}}}}}} & \begin{tabular}[c]{@{}l@{}}Free\\ express.\end{tabular} & \begin{tabular}[c]{@{}l@{}}Community members can freely express their thoughts and\\ opinions.\end{tabular}                                                              &\begin{tabular}[c]{@{}l@{}}High\end{tabular}                                                             \\ \hline
                                                                                     & Delib.                                              & \begin{tabular}[c]{@{}l@{}}Processes are in place that effectively and thoughtfully distill\\ options for decision-making from the collective intelligence.\end{tabular} & Medium  \\
\multirow{-3.5}{*}{\rotatebox[origin=c]{90}{\multirow{2}{*}{\textbf{\begin{tabular}[c]{@{}l@{}}Digital\\ Democ.\end{tabular}}}}}       & Voting                                                    & \begin{tabular}[c]{@{}l@{}}Choosing (desired) system states, goals and approaches to attain\\ them fairly from the options of the deliberation process.\end{tabular}      & Medium                                                 \\ \hline
                                                                                     & Autonomy                                                  & \begin{tabular}[c]{@{}l@{}}Community members can act as they think is best for\\ the organization.\end{tabular}                                                          & High \\
\parbox[t]{4mm}{\multirow{-2.8}{*}{\rotatebox[origin=c]{90}{\multirow{2}{*}{\textbf{\begin{tabular}[c]{@{}l@{}}Adap-\\ tation\end{tabular}}}}}}       & Feedback                                                  & (unsolicited real-time) feedback mechanisms are in place.                                                                                                                &High  \\ \hline                                                
\end{tabular}
\label{tab:dao_framework}
\end{table}

\section{Bitcoin's decentralized organization}
\label{sec:bitcoin_main_section}
Table \ref{tab:dao_framework} illustrates Bitcoins organizational mechanisms within the DAO viability framework. In the following, these are illustrated in greater detail.

\subsection{Collective Intelligence}

Bitcoin’s debate processes are intertwined across multiple online and offline platforms—ranging from specialized developer forums to mainstream social media—and offline venues such as conferences and local meetups. This mix ensures that different segments of the Bitcoin ecosystem can participate, collaborate, and influence the protocol’s evolution, all without the oversight of a centralized authority \cite{fish2024AnalyzingBitcoin}, which can facilitate collective intelligence to emerge \cite{helbing2023democracy}. In the following, the different collective intelligence principles are analyzed in greater detail.

\subsubsection{Openness}
Bitcoin has no legal entity that decides on the entry into the organization or expulsion of community members facilitating open boundaries to enter the organization, potentially being open to anyone wanting to contribute. Nevertheless, different roles within the organization exists, such as bitcoin holders, node operators or bitcoin core developer, which add (explicit) restrictions for entering or leaving. For instance, being a node operators implies having a particular technical skill set, and no formal processes exist are in place to apply or enter the Bitcoin core developer group.

Table \ref{tab:communication_channels} provides an overview of the different communication channels, virtual and physical, that the Bitcoin community uses to come together to discuss all aspects of Bitcoin - from the very technical protocol discussions to more educational fora for a wider group of interested people. None of the channels has formal barriers erected by the Bitcoin community.

\begin{table}[ht]
\centering
\begin{tabular}{p{4cm} p{2cm} p{7cm}}
\hline
\textbf{Channel / Event} & \textbf{Type} & \textbf{Key Characteristics} \\
\hline
bitcoin-dev mailing list, GitHub 
  & Virtual 
  & Debates on technical proposals and reviews (BIPs) \\ 
\hline
bitcointalk.org 
  & Virtual 
  & Original forum for all Bitcoin discourse in the early days. \\ 
\hline
\begin{tabular}[c]{@{}l@{}}Twitter/X, Telegram,  \\ Reddit, Discord \end{tabular} 
 
  & Virtual 
  & Dynamic platforms for community interaction and real-time updates \\ 
\hline
\begin{tabular}[c]{@{}l@{}}Specialized Channels  \\ (IRC, Slack, Mattermost) \end{tabular} 
  & Virtual 
  & Real-time discussions among developers and stakeholders \\ 
\hline
YouTube, Podcasts 
  & Virtual 
  & Interviews, tutorials, and capturing community sentiment \\ 
\hline
Onboarding programs 
  & Virtual 
  & Geared to onboard developers to Bitcoin core or related projects, eg bitcoindevs.xyz \\ 

\hline
Workshops
  & Physical 
  & Smaller developer gatherings around protocol upgrades (eg Scaling workshops) \\ 
\hline
Conferences 
  & Physical 
  & Organized larger-scale educational gatherings about economic, regulatory, industry topics  \\ 
\hline
Local Meetups 
  & Physical 
  & Knowledge sharing and networking within regional Bitcoin communities \\ 
\hline
    Coding Workshops, Hackathons 
  & Physical 
  & Hands-on collaboration and innovation for developers \\ 
\hline
University Seminars, Research Symposia 
  & Physical 
  & Academic rigor, research findings, and theoretical advancements in Bitcoin \\ 
\hline
\end{tabular}
\caption{Communication channels and events used by the Bitcoin community.}
\label{tab:communication_channels}

\end{table}

\subsubsection{Transparency}
Transparent access to the information concerning the organization are made available via various online and offline channels, as illustrated in Table \ref{tab:communication_channels}. 


The Bitcoin community employs online forums to collaboratively develop and discuss protocol modifications. Technical proposals and reviews primarily occur on the bitcoin-dev mailing list and GitHub, where Bitcoin Improvement Proposals (BIPs) (see Section \ref{sec:deliberation}) are introduced, debated, and refined. Long-established platforms such as Bitcointalk and Reddit facilitate broader discourse, covering everything from scaling debates to general news, while Twitter (X), Telegram, and Discord serve as more dynamic locations for community interaction and real-time updates. In parallel, specialized channels like IRC, Slack, or Mattermost host real-time discussions among developers and stakeholders, and YouTube/podcasts provide venues for interviews, tutorials, and community sentiment. 

Face-to-face engagements complement digital interaction by convening stakeholders in formal and informal settings. Prominent conferences (e.g., in Miami, Amsterdam, or El Salvador) bring together developers, miners, entrepreneurs, and investors to address protocol upgrades, regulatory shifts, and market dynamics. Local meetups in cities worldwide encourage knowledge sharing and networking, while coding workshops and hackathons offer opportunities for hands-on collaboration and innovation. Furthermore, university seminars and research symposia introduce academic rigor to Bitcoin’s development, enabling researchers to present findings, debate theoretical advancements, and shape the community’s technical direction. 

Collectively, these digital and physical spaces foster transparent participation, peer review, and consensus-building across diverse audiences.
In particular, suggestion to the bitcoin protocol are made early accessible to a broad audience.
The largest challenge might be to keep an overview and to know at which location to interact meaningfully with the community.

\subsubsection{Privacy}

No identity provision is required by Bitcoin, neither on-chain nor off-chain. Users of the bitcoin protocol can create new identities permissionless, nevertheless, these identities can be linked \cite{ballandies2022decrypting}. In particular, no inherent privacy mechanisms exist, making the exposure of identieis possible. 
Joining the digital discussion spaces usually requires an online identity that can also be created anew.
Also physical events usually do not require identification. In particular, there is no offical identity to be part of the Bitcoin organization or not.

\subsubsection{Freedom of expression}
\label{sec:freedom}
The Bitcoin community engages in open, public discussions that embrace a wide range of viewpoints and communication styles. The debate over Bitcoin maximalism—particularly regarding its toxicity or benefits —highlights this diversity, featuring opinions from support to opposition and discourse ranging from harsh disputes to neutral, nuanced exchanges \cite{boomer2012newcomer,buterin2022defense}.

\begin{table}[ht]
\centering
\begin{tabular}{p{1.8cm} p{2cm} p{2.2cm} p{1.8cm} p{4.2cm}}
\hline
\textbf{Princip.} & \textbf{Mechanism} & \textbf{Who} & \textbf{Locat.} & \textbf{How} \\
\hline
Deliberation 
 & \begin{tabular}[c]{@{}l@{}}Bitcoin \\ Improvement \\ Proposal \\ (BIP)\end{tabular} 
 & \begin{tabular}[c]{@{}l@{}}Core \\ Developers\end{tabular} 
 & \begin{tabular}[c]{@{}l@{}}Off-\\chain\end{tabular} 
 & Proposals are introduced, debated, and refined through a public and structured process \\ 
\hline
Voting 
 & \begin{tabular}[c]{@{}l@{}}Hash \\ Power \\ Voting\end{tabular} 
 & Miners 
 & \begin{tabular}[c]{@{}l@{}}On-\\chain\end{tabular} 
 & Miners signal support for changes by embedding markers (e.g., version bits) in mined blocks and eventually (not) upgrading their clients \\ 
\hline
Voting 
 & \begin{tabular}[c]{@{}l@{}}User-Activated \\ Soft Fork \\ (UASF)\end{tabular} 
 & \begin{tabular}[c]{@{}l@{}}Full \\ Nodes\end{tabular} 
 & \begin{tabular}[c]{@{}l@{}}On-\\chain\end{tabular} 
 & Full nodes enforce new rules by rejecting blocks that don’t comply after a specified date \\ 
\hline
\end{tabular}
\caption{Bitcoin's digital governance mechanisms.}
\label{tab:bitcoin-mechanisms}
\end{table}

\subsection{Digital Democracy: Decision-making}
The digital democracy component of a DAO consists of i) deliberation - methods to find voteable options and ii) voting - methods to agree within the community on these options \cite{ballandies2024daos}.
\subsubsection{Deliberation:}
\label{sec:deliberation}
Central to Bitcoin’s deliberation is the Bitcoin Improvement Proposal (BIP) process, moderated by the bitcoin core developers on github.
It is the only formal mechanism for its deliberation.
Stakeholders wishing to modify Bitcoin’s source code draft BIPs, which outline potential protocol changes, technical details, and implementation strategies. These proposals are typically reviewed by the wider development community on public forums (e.g., mailing lists, GitHub, see Table \ref{tab:communication_channels}) to solicit feedback. While Bitcoin Core developers maintain the reference implementation, they do not unilaterally decide on proposals; rather, extensive peer review and consensus-building shape the evolution of each BIP. Nevertheless, core developers have a veto right \cite{fish2024AnalyzingBitcoin}, as they ultimately decide on integrating code into the code base.


\subsubsection{Voting:}
Two voting mechanims can be identified within the Bitcoin community.

\textit{Hashpower voting} is executed by consensus participants ("miners") via their contributed hashpower:
After a new feature is integrated into Bitcoin Core, the network’s miners individually choose whether to upgrade. This opt-in model is crucial: miners show their preferences by installing or rejecting upgraded software. If a majority of the network adopts a particular change, it becomes the de facto rule set. Conversely, if the network fragments over a contentious proposal, a “fork” may emerge, leading to separate blockchains adhering to different consensus rules.
Several upgrade mechanisms have been utilized in the past by the miners  \cite{fish2024AnalyzingBitcoin}:
\begin{itemize}
    \item Flag day: All miners upgrade their nodes on a particular block height.
    \item BIP34 - Bitcoin miners indicate support for a BIP by adding a version bit to its produced block
    \item BIP9: followed BIP34 and introduced the ability for Miners to signal readiness for multiple upgrades with version bits in the block header
\end{itemize}

\textit{User voting: UASF and URSF}.
Beyond the miners, the so-called “economic majority”\footnote{https://en.bitcoin.it/wiki/Economic\_majority, last accessed: 2025-01-18} maintaining full nodes that independently from miners verify the blockchain's state —exchanges, payment processors, merchants, and large-scale holders—wields significant influence. These entities often shape decisions by signaling which version of Bitcoin software they support, since their adoption patterns can influence user confidence and liquidity. As a result, even if miners favor a change, that change may fail if influential economic actors do not upgrade or recognize the new rules, or vice versa, make miners upgrade to new rules that they otherwise would oppose. The former is referred to as User Resisted Soft Fork (URSF) and the latter as User Activated Soft Fork (UASF) \cite{fish2024AnalyzingBitcoin}. 


\subsection{Adaptation:}
The two principles of autonomy and feedback enable a DAO to adapt effectively to chaning environmental conditions \cite{ballandies2024daos}.
\subsubsection{Autonomy:}

Bitcoin has no legal entity or central authority such as a vocal founder that could dictate members of the organization how to act \cite{hsieh2018bitcoin}. Instead, Bitcoin is permissionless, meaning anyone can join and act for Bitcoin without permission. For instance, anyone is enabled to maintain a miner, hold bitcoins, or perform any other action on behalf of Bitcoin. In particular, any individual in the network decides in a case of a fork which chain is considered to be the main chain, which is ultimatively is decided by the collective action of these individuals. This voting by the feet characterizes also federal democracies \cite{frey1994swiss}. It is even argued by leading practitioners that this form of decision-making is the superior form when compared to token-based votings \cite{buterin2021moving}.

\subsubsection{Feedback:}
Blockchain-based tokens function as a feedback mechanisms facilitating self-organization in decentralized systems \cite{kleineberg2021social,dapp2021finance,ballandies2022incentivize}.
Bitcoin directly utilizes token units to incentivize the desireable behavior of mining. Also, the Bitcoin price can be seen as a further signal to the community on their overall performance. In particular, the recent stagnation of the Ethereum price is considered as an underperformance of its ecosystem, which initiated a change in the governance of the Ethereum foundation\footnote{https://cointelegraph.com/news/ethereum-foundation-infighting-and-drop-in-dapp-volumes-put-cloud-over-eth-price}.

Another feedback mechanism is the adoption of new software clients by miners \cite{fish2024AnalyzingBitcoin} and users, eventually providing feedback on which is the right implementation via the longest blockchain rule.


\section{Discussion} 
\label{sec:discussion}

A fundamental value of Bitcoin lies in its permissionless nature, exemplified by the absence of a legal entity or organization that owns it and could impose restrictions on the participation of community members. Given Bitcoin's success, this observation could serve as a starting point for discussions on whether DAOs should operate without legal entities in general. Specifically, the establishment of a legal entity can bind a DAO to a legal jurisdiction, potentially compromising its decentralized character and the autonomy of its participating members (Autonomy in Table \ref{tab:dao_framework}). However, adapting to changing environments in a decentralized manner relies on individual actors responding quickly to local conditions~\cite{ballandies2024daos}. Without a governing body capable of censoring, directing, or manipulating such responses, one could argue—based on Bitcoin's success—that this decentralized adaptability was instrumental in its growth and resilience.
However, recent work on DAO legal frameworks (e.g., DAO Model Law \cite{choi2021model}) reflects a discussion that DAOs must navigate legal ambiguity to engage in real-world operations. While Bitcoin’s governance is enabled by the absence of a legal entity—ensuring high autonomy—many DAOs benefit from limited liability structures or legal wrappers to interface with institutions.
This raises a tension: is the absence of a legal form a feature or a limitation? We argue that Bitcoin’s model expands the design space for DAOs by illustrating that meaningful coordination and value accrual can occur entirely outside of legal personhood. Nonetheless, future DAO designs may adopt hybrid forms, balancing legal interoperability with permissionless organizations.






Bitcoin utilizes a basic deliberation mechanism through its Bitcoin Improvement Proposals (BIPs) that has been adopted by many (smart-contract-based) DAOs. In contrast, digital democracy initiatives, such as those in Taiwan, have developed advanced deliberation tools that enhance reasoning, feedback, and sentiment analysis, thereby improving consensus-building and the identification of actionable options \cite{yang2023designing,helbing2023democracy}. However, considering the discussion on Bitcoin Maximalism (Section \ref{sec:freedom}), it remains questionable whether the Bitcoin community is willing to adopt new mechanisms. Nonetheless, introducing voluntary deliberation mechanisms in sub-communities of Bitcoin would not alter the underlying Bitcoin protocol; thus, a minority could initiate and experiment with them, potentially influencing the larger ecosystem if proven effective. 
According to the DAO viability framework, this would enhance the DAO's viability through more effective deliberation.

Also, how to vote and who the deciding parties are in Bitcoin is not clearly defined, which can raise confusion on who is the decisive party. In particular, the user-activated-soft-fork voting mechanism poses the risk of major economic actors unidirectionally changing Bitcoin. For instance, Blackrock is by now one of the largest Bitcoin holders. In their service agreement for their iShares trust, they specify that it is up to them to decide which Bitcoin chain will function as underlying to their ETFs in case of a Fork\footnote{"In the event of a hard fork of the Bitcoin network, the Sponsor [iShares Delaware Trust Sponsor LLC] will, if permitted by the terms of the Trust Agreement, use its discretion to determine which network should be considered the appropriate network for the Trust’s purposes, and in doing so may adversely affect the value of the Shares.", page 24, https://www.sec.gov/Archives/edgar/data/1980994/000143774923028549/bit20231017\_s1a.htm, last accessed: 2025-01-21}. In case, Blackrock and some other major actors like Microstrategy decide to go for a chain that does not align with the larger Bitcoin ecosystem, this could pose a significant risk to the value of Bitcoin and subsequently its security \cite{ballandies2022decrypting}. In particular, there is a signficant risk that such major economic actors support values that do not align with the larger community of Bitcoin holders.

\subsection{Limitations}
\label{sec:limitations}
Elinor Ostroms design principles for successfully governing the commons \cite{ostrom1990governing} have been applied to illustrate DAOs \cite{hunhevicz2024decentralized}. Taking such a commons view on DAOs, the DAO viability framework could be extended with those principles to illustrate in how far a DAO is viable to govern a commons. In particular, it would provide a more differentiated view into the Openess aspect of collective intelligence (Table \ref{tab:dao_framework}), as a commons needs to have clear boundaries and sanctioning mechanisms defined \cite{ostrom1990governing}. Bitcoin seems to fulfill this differentiated view on the boundaries, as people can enter and leave freely the organization, but entering the inner circles is less transmissive (e.g. becoming a core developer). 

Also, the performed literature analysis could be extended to involve further scientific databases. In particular, investigating the sentiment towards Bitcoin being a DAO in social-science related databases would provide a greater nuanced view on the perception of Bitcoin being a DAO, as a significant amount of research on DAOs is performed in these fields \cite{parmindar2025conceptual}.
Moreover, the quantitative analysis could be strengthened by adding further sentiment analysis methods besides ChatGPT. 

\section{Conclusion and Outlook}
\label{sec:conclusion}
This work demonstrates that Bitcoin can be regarded as a decentralized autonomous organization, instantiating the three self-organization mechanisms of the DAO viability framework—adaptation, digital democracy, and collective intelligence (Table \ref{tab:dao_framework})—through innovative methods such as decentralized node signaling for voting. Given its economic success and the ongoing governance debate within the Bitcoin community, the DAO research community would benefit from analyzing Bitcoin's DAO mechanisms to derive insights into successful DAO instantiation. In turn, practitioners could draw on these scholarly insights to inform their discussions.

For instance, applying the DAO viability framework, we identified several mechanisms supporting Bitcoin's viability—such as the absence of a legal entity—that stem from its core value of permissionlessness. We also found that Bitcoin governance could be enhanced by improving deliberation, potentially initiated by a sub-group within the ecosystem integrating recent findings from digital democracy research.

In general, Bitcoin was described early on as a decentralized corporation exhibiting core DAO features such as open participation and rule enforcement via code. That the term "DAO" emerged later does not diminish Bitcoin’s alignment with the concept—in fact, Bitcoin helped shape it. While smart contract-based DAOs emphasize programmable governance, Bitcoin represents an alternative archetype of a DAO focused on social consensus and permissionless actions. Recognizing this divergence broadens the DAO design space and affirms Bitcoin's continued relevance as a living, functional DAO.

Beyond refining the DAO viability framework through Ostrom’s work (Section \ref{sec:limitations}), two further research directions emerge. First, Bitcoin’s on-chain miner signaling remains understudied; network science could reveal key governance dynamics. Second, quantitative methods could assess adherence to the eight viability principles, potentially using machine learning or network analysis methodologies. This would lay the foundation for quantitatively accessing the viability of a DAO.

\appendix
\section{ChatGPT Prompt}
\label{sec:chatgp_prompt}

\begin{lstlisting}
PROMPT_TEMPLATE = """As an expert in blockchain research, you are analyzing a paper mentioning Bitcoin and Decentralized Autonomous Organizations (DAOs).

Your task is to:
1. Determine whether the paper supports or opposes the idea that "Bitcoin is a DAO."
2. Determine the author's overall sentiment toward 'Bitcoin is a DAO', give true or false. In addition, give a sentiment score.
3. Give a general sentiment score. If the author likes bitcoin or think bitcoin has great potential, the score should be high.
3. Provide supporting reasoning for 'bitcoin is a DAO', including cases where both supporting and opposing arguments are present.

Content:
{text}

Output requirements:
Format: JSON
Fields:
"dao_relation": A boolean (true or false) indicating whether the author ultimately supports (true) or opposes (false) the idea that Bitcoin is a DAO.
"sentiment_score": A float between 0 and 1, representing the author's overall sentiment toward 'Bitcoin is a DAO' (0 = very negative, 1 = very positive).
"general_sentiment_score": A float between 0 and 1, representing the author's overall sentiment toward bitcoin (0 = very negative, 1 = very positive).
"reasoning": A concise summary of the author's stance, highlighting supporting and opposing arguments if both exist.
"citations": A list of locations where relevant arguments appear

Example output:
{{"dao_relation": true, "sentiment_score": 0.85, "general_sentiment_score": 0.4, "reasoning": "The author explicitly states that 'bitcoin is a DAO' in paragraph 3. "citation": ["paragraph 3", "bitcoin can be considered a DAO due to its decentralized nature"]}}

Important Notes:
If the paper presents both supporting and opposing arguments, determine whether the overall stance leans more towards supporting (true) or opposing (false) and classify "dao_relation" accordingly.
If mixed arguments exist, explicitly mention both views in "reasoning" and cite them in "citations".
"""
\end{lstlisting}

\section*{Acknowledgements}
We thank Marcus M. Dapp for his valuable feedback on this manuscript.

 \bibliographystyle{styles/bibtex/splncs03}
 \bibliography{paper}

\end{document}